\documentclass[a4paper, onecolumn]{article}
\usepackage{setspace}
\usepackage{textcomp}
\usepackage{diagbox}
\usepackage{titlesec}
\usepackage[titletoc,toc,page,header]{appendix}
\usepackage{tabularx}
\usepackage{authblk}
\usepackage{booktabs}
\usepackage{graphicx}
\usepackage{amsmath}
\usepackage{geometry}
\usepackage{amssymb}
\usepackage{float}
\usepackage{makecell}
\usepackage{multirow}
\usepackage{tikz}
\usepackage{cite}
\usepackage{gensymb}
\usepackage{amstext}
\usepackage{bm}
\usepackage{indentfirst}
\usepackage{latexsym}
\usepackage{color}
\usepackage{subfigure}
\usepackage{slashed}
\usepackage{hyperref}
\usepackage{multicol}
\usepackage{lipsum}
\usepackage{widetext}
\usepackage{cuted}
\usepackage{flushend}
\usepackage{threeparttable}
\usetikzlibrary{trees}
\usetikzlibrary{decorations.pathmorphing}
\usetikzlibrary{decorations.markings}
\usepackage{mathrsfs}
\usepackage{amsbsy}
\tikzset{
   global scale/.style={
      scale=#1,
      every node/.append style={scale=#1}},
   photon/.style={decorate, decoration={snake}, draw=red},
   nucleon/.style={draw=black, postaction={decorate},
      decoration={markings,mark=at position .55 with{\arrow[draw=black]{>}}}},
   pion/.style={draw=blue, postaction={decorate},
      decoration={markings,mark=at position .55 with{\arrow[draw=blue]{}}}},
    }

\newcommand{\be}{\begin{equation}} \newcommand{\ee}{\end{equation}}
\newcommand{\ba}{\begin{array}{c}} \newcommand{\ea}{\end{array}}
\newcommand{\bea}{\begin{eqnarray}} \newcommand{\eea}{\end{eqnarray}}

\allowdisplaybreaks{}

\begin{document}
\title{\Large\textbf{Dispersive Analysis of Low Energy \(\gamma N\rightarrow\pi N\) Process and Studies on the \(N^*(890)\) Resonance}}
\author{Yao Ma\(^1\)}
\author{Wen-Qi Niu\(^1\)}
\author{De-Liang Yao\(^2\)}
\author{Han-Qing Zheng\(^{1,3}\)}
\affil{\(^1\)Department of Physics and State Key Laboratory of Nuclear Physics and Technology, \\Peking University, Beijing 100871, P. R. China}
\affil{\(^2\)School of Physics and Electronics, Hunan University, Changsha 410082, P. R. China}
\affil{\(^3\)Collaborative Innovation Center of Quantum Matter, Beijing, Peoples Republic of China}

\maketitle

\begin{abstract}
We present a dispersive representation of the \(\gamma N\rightarrow \pi N\) partial-wave amplitude based on unitarity and analyticity.
In this representation, the right-hand-cut contribution responsible for \(\pi N\) final-state-interaction effects are taken into account via an Omn\'es formalism with elastic \(\pi N\) phase shifts as inputs, while the left-hand-cut contribution is estimated by invoking chiral perturbation theory.
Numerical fits are performed in order to pin down the involved subtraction constants.
It is found that good fit quality can be achieved with only one free parameter and the experimental data of the multipole amplitude \(E_{0}^+\) in the energy region below the \(\Delta(1232)\) are well described.
Furthermore, we extend the \(\gamma N\rightarrow \pi N\) partial-wave amplitude to the second Riemann sheet so as to extract the couplings of the \(N^\ast(890)\).
The modulus of the residue of the multipole amplitude \(E_{0}^+\) (\({\rm S_{11}pE}\)) is \(2.41\rm{mfm\cdot GeV^2}\) and the partial width of \(N^*(890)\to\gamma N\) at the pole is about \(0.369\ {\rm MeV}\), which is almost the same as the one of the \(N^*(1535)\) resonance, indicating that \(N^\ast(890)\) strongly couples to \(\pi N\) system.
\end{abstract}
%\keywords{photoproduction, \(N^*(890)\), residue}

\section{Introduction}
Single pion photoproduction off the nucleon has been extensively studied for its importance in determining the spectrum and properties of the nucleon resonances~\cite{Chew1957,Adler:1968tw,Review1969,Drechsel:2007if}.
There have been many measurements on this process, accumulating a wealth of experimental data on, e.g., cross section, photon asymmetry, target asymmetry, etc; see e.g. Refs~\cite{Benz:1974tt,Fuchs:1996ja,Blanpied:2001ae,Ahrens:2004pf,SAID}.
Based on the dataset, partial wave analyses were performed to anatomize the underlying structure of the reaction amplitude and justify the existence of the nucleon resonances theoretically.
At low energies, it has been successful to explore the photoproduction processes in chiral perturbation theory (ChPT)~\cite{Bernard:1991rt,Bernard:1992nc,Bernard:2001gz,Hilt:2013uf,Hilt:2013fda,Blin:2014rpa,Blin:2016itn,Navarro:2019iqj}.
In combination of unitarization approaches\cite{Martin1976}, the valid region of the chiral amplitudes is extended and physical states behave themselves as pole singularities of the unitarized amplitudes.
Nevertheless, most of the unitarization methods only take the unitary cut into account, while the remaining left-hand cuts (l.h.c.s) are left out, leading to the drawback that the proper analytic and crossing properties of the amplitude is not faithfully guaranteed.
In consequence, spurious poles arise to mimic the contribution of the left-hand cuts, or even worse, prevent us from discovering certain truly existent poles, e.g.\ virtual poles or subthreshold resonances.

In Refs.~\cite{Wang2018,Wang2019,Wang2018a}, a novel subthreshold resonance named \(N^\ast(890)\) was found in the \(S_{11}\) wave through a prudent analysis of the covariant chiral amplitude of \(\pi N\) scattering~\cite{Chen:2012nx,Alarcon:2012kn,Yao:2016vbz,Siemens:2017opr} by applying the method of Peking University (PKU) representation~\cite{Xiao:2000kx,He:2002ut,Zheng:2003rw,Zheng:2003rv,Zhou:2004ms,Zhou:2006wm}.
The PKU representation respects causality and has already been used to establish the existences of the \(\sigma \) and \(\kappa \) states~\cite{Xiao:2000kx,Zheng:2003rw}. The discovery of the \(N^\ast(890)\) resonance is nothing but an improved implement of analyticity compared to other unitarization methods. For instance, it is pointed out in Ref.~\cite{Ma:2020sym} that the \(N^\ast(890)\) resonance still exists even in a \(K\)-matrix parametrization if a better treatment of analyticity is executed. {However, it should be emphasized that, in the traditional $K$ matrix method without any improvement of analyticity, even if a pole emerges from the background polynomial, it is not legitimate to discuss whether it is physical or not, it only means the non-background part of the $K$ matrix parametrization is incomplete in characterizing whole physics. In PKU representation, the existence of $N^\ast(890)$ actually only depends on our understanding or knowledge of left hand cut contribution at qualitative level -- that is its contribution to the phase shift is negative. }
In this paper, we intend to explore the \(N^\ast(890)\) resonance in the \(\gamma N\rightarrow \pi N\) scattering to gain more information on its properties.

Our \(\gamma N\rightarrow \pi N\) amplitudes are obtained through a dispersive representation, which is set up with the help of unitarity and analyticity\cite{Babelon1976,Babelon1976a,Mao2009,Dai:2016ytz}.
The inputs of the dispersive representation are \(\pi N\) final-state-interaction amplitude and chiral tree-level \(\gamma N\rightarrow \pi N\) amplitude estimating the left hand singularities of pion photoproduction.
In a single channel approximation, the former can be achieved by an Omn\`{e}s solution with the \(\pi N\) scattering phase as input.
The left hand cuts are calculated based on a chiral Lagrangian with pion and nucleon fields truncated at order \(q^2\).
We review the analytic structures of pion photoproduction amplitudes in Ref.~\cite{Kennedy1962} and analyze the relevant singularities to arising in our calculation.
In addition, we find that kinematical singularities in this inelastic process are rather complicated. Cuts coming from kinematical structure depend on how to organize the analytic functions in the amplitudes.
These cuts could be in the complex plane and may affect the residues of \(N^*(890)\). To avoid such complexity, we deform these cuts in a particular way to make sure that they are lying on the real axis, below the pseudo threshold of \(\pi N\) scattering.

We fit the multipole amplitudes \(E_{0+}\) (\(\rm S_{11}pE\) and \(\rm S_{11}nE\)) from Ref.~\cite{Workman2012} below the \(\Delta(1232)\) peak in order to determine the subtraction polynomial in the dispersive representation. The residue couplings of \(N^*(890)\) can be computed by analytic continuation of the amplitude to second sheet, in which the PKU representation of \(\pi N\) \(\mathcal{S}\) matrix is employed.
We compare the residues of \(N^*(890)\) extracted from multipole amplitudes with the ones of \(N^*(1535)\) obtained in Ref.~\cite{Svarc2014} to learn the properties of \(N^*(890)\) and get some information of structures by analogy with the analysis of \(N^*(1535)\).

The structure of this paper is organized as follows.
In Section~\ref{sec:df} we set up the dispersive formalism for \(\gamma N \to \pi N\) process.
Then the left-hand-cuts are estimated based on chiral perturbation theory in Section~\ref{sec:lhcchpt}, and we also make an analysis about the singularities which will appear in this pion photoproduction process.
In last two sections, numerical results are carried out and summary is presented.

\section{Dispersive formalism for \texorpdfstring{\(\gamma N\to\pi N\)}{Lg}}\label{sec:df}

\subsection{Dispersive representation}
The unitarity relation for the \(\gamma N\to \pi N^\prime \) partial wave amplitude reads
\begin{equation}\label{eq:unitarity}
    \frac{\mathcal{M}(s+i\epsilon)-\mathcal{M}(s-i\epsilon)}{2i}={\rm Im}\mathcal{M}(s+i\epsilon)= \mathcal{T}^*(s+i\epsilon)\rho(s+i\epsilon)\mathcal{M}(s+i\epsilon)\ ,
\end{equation}
where \(\mathcal{T}\) is pion-nucleon scattering amplitude in \(S_{11}\) wave.
The function \(\rho(s)\) is defined by
\begin{eqnarray}
    \rho(s)=\frac{\sqrt{(s-s_L)(s-s_R)}}{s}\ ,
\end{eqnarray}
where \(s_{R}\equiv{(m_N+m_\pi)}^2\) and \(s_L\equiv{(m_N-m_\pi)}^2\).
Equivalently, Eq.~\eqref{eq:unitarity} can be recast to
\begin{eqnarray}\label{eq:uni2}
    \mathcal{M}(s+i\epsilon)=\mathcal{S}(s+i\epsilon)\mathcal{M}(s-i\epsilon)\ ,
\end{eqnarray}
where \(\mathcal{S}(s)=1+2i\rho(s)\mathcal{T}(s)\) which is the \(\pi N\) scattering \(S\) matrix in single channel case.
The scattering amplitude \(\mathcal{M}\) may be separated into two parts, i.e., \(\mathcal{M}=\mathcal{M}_R+\mathcal{M}_L\).
The former part \(\mathcal{M}_R\) only contains the right hand cut (RHC) starting at \(s_R\), while the latter part \(\mathcal{M}_L\) is free of RHC singularity.
Substituting \(\mathcal{M}=\mathcal{M}_R+\mathcal{M}_L\) into Eq.~\eqref{eq:uni2}, one gets
\begin{eqnarray}\label{eq:uni3}
    \mathcal{M}_R^+=\mathcal{S}\mathcal{M}_R^-+(\mathcal{S}-1)\mathcal{M}_L\ .
\end{eqnarray}
For convenience, the abbreviations \(\mathcal{M}^{\pm}(s)=\underset{\epsilon\to0}{\lim}\mathcal{M}(s\pm i\epsilon)\) have been used.
To proceed, we introduce a helper function \(\mathcal{D}(s)\) which is analytic throughout the complex \(s\) plane but encodes the same unitarity singularity as \(\mathcal{M}(s)\).
Namely, it satisfies the same unitarity condition as \(\mathcal{M}(s)\) along the unitary cut:
\begin{eqnarray}\label{eq:uni4}
    \frac{\mathcal{D}^+}{\mathcal{D}^-}=\frac{\mathcal{M}^{+}}{\mathcal{M}^{-}}=\mathcal{S}\ .
\end{eqnarray}
By expressing the \(\mathcal{S}\) matrix in Eq.~\eqref{eq:uni3} by \(\mathcal{D}(s)\), the following relation of spectral functions can be obtained:
\begin{eqnarray}\label{eq:rhue}
    {\rm Im}\big(\mathcal{D}^{-1}\mathcal{M}_R\big) = -\big({\rm Im}\mathcal{D}^{-1}\big)\mathcal{M}_L \ .
\end{eqnarray}
Straightforwardly, a dispersive representation for \(\mathcal{M}_R\) can be written down,
\begin{equation}\label{drpwa}
    \mathcal{M}_R(s)=\mathcal{D}\left(-\frac{s^n}{\pi}\int_{s_R}^{\infty}\frac{\big({\rm Im} \mathcal{D}^{-1}\big)\mathcal{M}_L}{s'^n(s'-s)}{\rm d} s'+\mathcal{P}\right)\ ,
\end{equation}
where \(n\) is the number of subtractions and \(\mathcal{P}(s)\) is subtraction polynomial.
Eventually,
\begin{eqnarray}\label{eq:disRep}
    \mathcal{M}(s)=\mathcal{M}_L+\mathcal{D}\left(-\frac{s^n}{\pi}\int_{s_R}^{\infty}\frac{\big({\rm Im} \mathcal{D}^{-1}\big)\mathcal{M}_L}{s'^n(s'-s)}{\rm d} s'+\mathcal{P}\right)\ .
\end{eqnarray}
Thus, the pion photoproduction amplitude \(\mathcal{M}(s)\) is determined up to a polynomial, once \(\mathcal{D}(s)\) and \(\mathcal{M}_L(s)\) are known.

Based on the unitarity condition in Eq.~\eqref{eq:uni4}, one can write a spectral representation for the auxiliary function \(\mathcal{D}(s)\),
\begin{eqnarray}
    \mathcal{D}(s)=\frac{1}{\pi}\int_{s_R}^\infty\frac{\mathcal{T}^\ast(s)\rho(s^\prime)\mathcal{D}(s^\prime)}{s^\prime-s}{\rm d}s^\prime \ .
\end{eqnarray}
The above representation yields an integral equation for \(\mathcal{D}(s)\), which has the so-called Omn\'es solution~\cite{Omnes:1958hv}
\begin{eqnarray}\label{eq:omnes}
    \mathcal{D}(s)=\tilde{\mathcal{P}}(s)\exp\bigg[\frac{s}{\pi}\int_{s_R}^\infty\frac{\delta(s^\prime)}{s^\prime(s^\prime-s)}{\rm d}s^\prime\bigg]
\end{eqnarray}
with \(\tilde{\mathcal{P}}\) standing for zero points in complex plane and \(\delta(s)\) being the elastic \(\pi N\) phase shift, in accordance with the Watson final state interaction (FSI) theorem~\cite{Watson:1954uc}.

\section{Estimate on the left-hand-cut contribution in ChPT}\label{sec:lhcchpt}
\subsection{Basics of single one-pion photoproduction off the nucleon}
Single one-pion photoproduction off the nucleon (\(\gamma N\)-\(1\pi \)) is the process as described by
\begin{eqnarray}
    \gamma(q)+N(p)\to \pi^a(q^\prime)+N^\prime(p^\prime)\ ,
\end{eqnarray}
where \(a\) is the isospin index of the pion and momenta of the particles are indicated in the parentheses. The isospin structure of the scattering amplitude can be written as
\begin{eqnarray}
    \mathcal{M}(\gamma+N\to \pi^a+N^\prime)=\chi_N^\prime \bigg \{ \delta_{a3}\,\mathcal{M}^++\frac{1}{2}[\tau_a,\tau_3]\,\mathcal{M}^-+\tau_3\,\mathcal{M}^0\bigg \} \chi_N \ ,
\end{eqnarray}
where \(\tau_a\) (\(a=1,2,3\)) are Pauli matrices in isospin space. Amplitudes with definite isospin \(I=\frac{1}{2},\frac{3}{2}\) can be obtained from \(\mathcal{M}^{\pm}\) and \(\mathcal{M}^0\) via
\footnote{The convention in Ref.~\cite{PDG2018} is adopted for the CG coefficients and for the physical pion states we use \( \pi^{+}=-\frac{1}{\sqrt{2}}\left(\pi_1-i\pi_2\right)\) and \(\pi^{-}=\frac{1}{\sqrt{2}}\left(\pi_1+i\pi_2\right)\).}
\footnote{\(\mathcal{M}\) and \(\mathcal{P}\) are actually vectors with two components in isospin space of \(I=\frac{1}{2}\) channel due to target asymmetry caused by electromagnetic interaction.}
\begin{align}
    \mathcal{M}^{I=\frac{3}{2}}=&\sqrt{\frac{2}{3}}\left(\mathcal{M}^{+}-\mathcal{M}^{-}\right)\ ,\\
    \mathcal{M}^{I=\frac{1}{2}}=&-\frac{1}{\sqrt{3}}\left(\mathcal{M}^{+}+2\mathcal{M}^{-}+3\mathcal{M}^{0}\right)\ , \  \text{(\(p\) target)}\\
    \mathcal{M}^{I=\frac{1}{2}}=&\frac{1}{\sqrt{3}}\left(\mathcal{M}^{+}+2\mathcal{M}^{-}-3\mathcal{M}^{0}\right) \ , \  \text{(\(n\) target)}\ .
\end{align}

The isospin amplitudes \(\mathcal{M}^{I}\) with either \(I=\frac{1}{2},\frac{3}{2}\) or \(I=\pm,0\) can be further decomposed in terms of four independent Lorentz operators as
\begin{equation}\label{eq:LorAmp}
    \mathcal{M}^{I}(s,t)\equiv\bar{u}(p^\prime)\mathcal{T}^{I}u(p)=\bar{u}(p^\prime)\bigg[\sum_{i=1}^{4}\mathcal{A}_i^{I}(s,t)\,{L}^i_\mu \epsilon^\mu \bigg]u(p)\ ,
\end{equation}
where
\begin{align}\label{ltd}
    L^1_{\mu}=&i\gamma_5\gamma_{\mu}\gamma\cdot q\ ,\nonumber \\
    L^2_{\mu}=&2i\gamma_5\left(P_{\mu}q\cdot q'-q'_{\mu}P\cdot q\right)\ ,\nonumber \\
    L^2_{\mu}=&\gamma_5\left(\gamma_{\mu}q'\cdot q-q'_{\mu}\gamma\cdot q\right)\ ,\nonumber \\
    L^4_{\mu}=&2\gamma_5\left(\gamma_{\mu}P\cdot q-P_{\mu}\gamma\cdot q\right)\ .
\end{align}
Note that the operators \(L^i_\mu \) obey the Ward identity~\cite{Chew1957}. 
Here \(\epsilon_\mu \) is the polarization vector of the photon, \(u(p)\) and \(\bar{u}(p^\prime)\) are the spinors of the nucleons.

\subsection{Calculation of chiral amplitudes at tree level}\label{lddabcd}
The effective Lagrangian for our calculation of the chiral amplitude up to \(\mathcal{O}(p^2)\) reads
\begin{eqnarray}
\mathcal{L}_{\rm eff} = \mathcal{L}_{\pi N}^{(1)}+\mathcal{L}_{\pi N}^{(2)}+\mathcal{L}_{\pi \pi}^{(2)}
\end{eqnarray}
with the superscripts referring to chiral orders. The terms in the above equation are given by~\cite{Scherer2012}
\begin{align}\label{eq:lag}
    \mathcal{L}_{\pi N}^{(1)}=&\bar{\Psi}\left(i\slashed{D}-m+\frac{g}{2}\gamma^{\mu}\gamma_5u_{\mu}\right)\Psi \ ,\\
    \mathcal{L}_{\pi N}^{(2)}=&\bar{\Psi}\sigma^{\mu\nu}\left[\frac{c_6}{2}f^+_{\mu\nu}+\frac{c_7}{2}v_{s,\mu\nu}\right]\Psi\ ,\nonumber \\
    \mathcal{L}_{\pi\pi}^{(2)}=&\frac{F^2}{4}Tr\left[D_{\mu}U{\left(D^{\mu}U\right)}^{\dagger}\right]+\frac{F^2}{4}Tr\left(\chi U^{\dagger}+U\chi^{\dagger}\right)\ ,
\end{align}
where \(m\), \(g\) and \(F\) are nucleon mass, nucleon axial coupling and pion decay constant in the chiral limit, in order.
Given our working accuracy, they are set equal to their physical counterparts, \(m_N\), \(g_A\) and \(F_\pi \), the physical nucleon mass, physical axial charge and pion decay constants.
Namely, \(m=m_N\), \(g=g_A\) and \(F=F_\pi \). Here \(c_{6}\) and \(c_7\) are \(\mathcal{O}(p^2)\) low energy constants (LECs) which are known parameters to be determined by experimental data;
See Ref.~\cite{Scherer2012} for definitions of the chiral blocks.

\begin{figure}[htb]\centering
    \subfigure[\(s\) channel]{
        \begin{tikzpicture}[global scale = 0.5]
            \draw[photon, very thick](-1.5,1.5)node[above]{\(q,\lambda\)}--(0,0);
%            \draw[->](-1,2)--(0,1);
            \draw[nucleon, very thick](-1.5,-1.5)node[below]{\(p,s\)}--(0,0);
            \draw[nucleon, very thick](0,0)--(2,0);
            \draw[nucleon, very thick](2,0)--(3.5,-1.5)node[below]{\(p',s'\)};
            \draw[pion, dashed, very thick](2,0)--(3.5,1.5)node[above]{\(q',a'\)};
%            \draw[->](2,1)--(3,2);
            \draw[very thick](0,0)circle(0.4);
            \draw[very thick](2,0)circle(0.4);
            \fill[white](0,0) circle(0.4);
            \node{\(2\)};
            \fill[white](2,0) circle(0.4);
            \node at (2,0) {\(1\)};
        \end{tikzpicture}
    }
    \subfigure[\(u\) channel]{
        \begin{tikzpicture}[global scale = 0.5]
            \draw[photon, very thick](-1.5,1.5)node[above]{\(q,\lambda\)}--(2,0);
%            \draw[->](-1,2)--(-0.3,1.7);
            \draw[nucleon, very thick](-1.5,-1.5)node[below]{\(p,s\)}--(0,0);
            \draw[nucleon, very thick](0,0)--(2,0);
            \draw[nucleon, very thick](2,0)--(3.5,-1.5)node[below]{\(p',s'\)};
            \draw[pion, dashed, very thick](0,0)--(3.5,1.5)node[above]{\(q',a'\)};
%            \draw[->](2,1.5)--(2.7,1.8);
            \draw[very thick](0,0)circle(0.4);
            \draw[very thick](2,0)circle(0.4);
            \fill[white](0,0) circle(0.4);
            \node{\(1\)};
            \fill[white](2,0) circle(0.4);
            \node at (2,0) {\(2\)};
        \end{tikzpicture}
    }
    \caption{\(\mathcal{O}(p^2)\) diagram}\label{trfd2}
\end{figure}
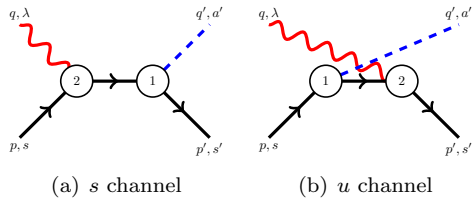

The relevant pieces extracted from the expanded form of the Lagrangians in Eq.~\eqref{eq:lag} are
\begin{align}
    \mathcal{L}^{(1)}_{\pi N}\supset&\,+\frac{g_A}{2F_{\pi}}\partial_{\mu}\phi\bar{\Psi}\gamma_5\gamma^{\mu}\Psi -\frac{e}{2}A_{\mu}\bar{\Psi}\left[\gamma^{\mu}\left(\tau_3+1\right)\right]\Psi -i\frac{eg_A}{4F_{\pi}}A_{\mu}\bar{\Psi}\left(\gamma_5\gamma^{\mu}\left[\phi,\tau_3\right]\right)\Psi\ , \\
    \mathcal{L}^{(2)}_{\pi N}\supset&\,-e\bar{\Psi}\sigma^{\mu\nu}\left[\frac{c_6}{2}\left(\partial_{\mu}A_{\nu}-\partial_{\nu}A_{\mu}\right)\tau_3+\frac{c_7}{4}\left(\partial_{\mu}A_{\nu}-\partial_{\nu}A_{\mu}\right)\right]\Psi\ ,\\
    \mathcal{L}_{\pi\pi}^{(2)}\supset&-\frac{ie}{8}A^{\mu}Tr\left( \left \{ \partial_{\mu}\phi,\left[\phi,\tau_3\right] \right \} \right)\ .
\end{align}
Tree-level Feynman diagrams up to \(\mathcal{O}(q^2)\) are displayed in Figures~\ref{trfd1} and~\ref{trfd2}.

\begin{figure}[H]\centering
    \subfigure[\(s\) channel]{
        \begin{tikzpicture}[global scale = 0.5]
            \draw[photon, very thick](0,0)to(-1.5,1.5)node[above]{\(q,\lambda\)};
            \draw[nucleon, very thick](-1.5,-1.5)node[below]{\(p\)}to(0,0);
            \draw[nucleon, very thick](0,0)to(2.5,0);
            \draw[nucleon, very thick](2.5,0)to(4,-1.5)node[below]{\(p'\)};
            \draw[pion, dashed, very thick](2.5,0)to(4,1.5)node[above]{\(q',a\)};
            \draw[very thick](0,0)circle(0.4);
            \draw[very thick](2.5,0)circle(0.4);
            \fill[white](0,0)circle(0.4);
            \fill[white](2.5,0)circle(0.4);
            \node at(0,0){\(1\)};
            \node at(2.5,0){\(1\)};
        \end{tikzpicture}
    }
    \subfigure[\(u\) channel]{
        \begin{tikzpicture}[global scale = 0.5]
            \draw[photon, very thick](2.5,0)to(-1.5,1.5)node[above]{\(q,\lambda\)};
            \draw[nucleon, very thick](-1.5,-1.5)node[below]{\(p\)}to(0,0);
            \draw[nucleon, very thick](0,0)to(2.5,0);
            \draw[nucleon, very thick](2.5,0)to(4,-1.5)node[below]{\(p'\)};
            \draw[pion, dashed, very thick](0,0)to(4,1.5)node[above]{\(q',a\)};
            \draw[very thick](0,0)circle(0.4);
            \draw[very thick](2.5,0)circle(0.4);
            \fill[white](0,0)circle(0.4);
            \fill[white](2.5,0)circle(0.4);
            \node at(0,0){\(1\)};
            \node at(2.5,0){\(1\)};
        \end{tikzpicture}
    }
    \subfigure[contact diagram]{
        \begin{tikzpicture}[global scale = 0.5]
            \draw[pion, dashed, very thick](0,0)--(1.5,1.5)node[above]{\(q',a\)};
            \draw[photon, very thick](0,0)--(-1.5,1.5)node[above]{\(q,\lambda\)};
            \draw[nucleon, very thick](0,0)--(1.5,-1.5)node[below]{\(p'\)};
            \draw[nucleon, very thick](-1.5,-1.5)node[below]{\(p\)}--(0,0);
            \draw[very thick](0,0) circle (0.4);
            \fill[white](0,0) circle (0.4);
            \node at(0,0){\(1\)};
        \end{tikzpicture}
    }
    \subfigure[\(t\) channel]{
        \begin{tikzpicture}[global scale = 0.5]
            \draw[photon, very thick](-1.5,1.5)node[above]{\(q,\lambda\)}--(0,0);
            \draw[pion, dashed, very thick](0,0)--(1.5,1.5)node[above]{\(q',a\)};
            \draw[pion, dashed, very thick](0,0)--(0,-2.5);
            \draw[nucleon, very thick](-1.5,-4)node[below]{\(p,s\)}--(0,-2.5);
            \draw[nucleon, very thick](0,-2.5)--(1.5,-4)node[below]{\(p',s'\)};
            \fill[white](0,0) circle(0.4);
            \node(0,0){\(2\)};
            \fill[white](0,-2.5) circle(0.4);
            \node at(0,-2.5){\(1\)};
            \draw[very thick](0,0) circle(0.4);
            \draw[very thick](0,-2.5) circle(0.4);
        \end{tikzpicture}
    }
    \caption{\(\mathcal{O}(p)\) diagram}\label{trfd1}
\end{figure}
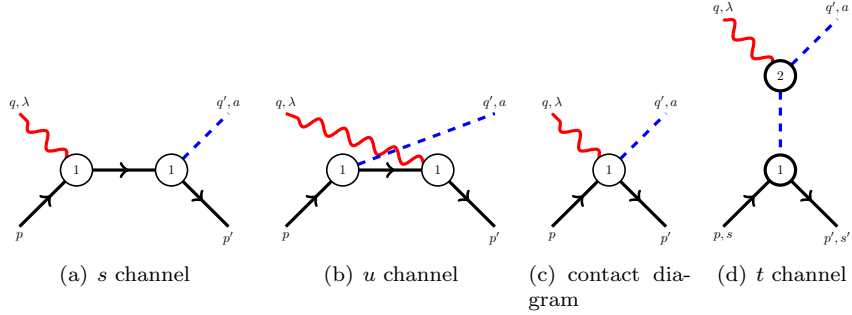

The full amplitude reads
\begin{align}
    i\mathcal{M}^{(1)}=&\frac{eg_A}{4F_{\pi}}\chi^{\dagger}_f\left[\tau_a,\tau_3\right]\chi_i\bar{u}_{s'}(p')\gamma_5\gamma^{\mu}u_s(p)\epsilon_{\lambda,\mu}(q)\nonumber \\
    &+\frac{ieg_A}{4F_{\pi}}\chi^{\dagger}_f\tau_a\left(\tau_3+1\right)\chi_i\bar{u}_{s'}(p')\gamma_5\gamma^{\nu}\frac{i}{\slashed{p}+\slashed{q}-m_N+i\epsilon}\gamma^{\mu}u_s(p)q'_{\nu}\epsilon_{\lambda,\mu}(q)\nonumber \\
    &+\frac{ieg_A}{4F_{\pi}}\chi^{\dagger}_f\left(\tau_3+1\right)\tau_a\chi_i\bar{u}_{s'}(p')\gamma^{\mu}\frac{i}{\slashed{p}'-\slashed{q}-m_N+i\epsilon}\gamma_5\gamma^{\nu}u_s(p)q'_{\nu}\epsilon_{\lambda,\mu}(q)\nonumber \\
    &-\frac{ieg_{A}}{4F_{\pi}}\chi_f^{\dagger}\left[\tau_a,\tau_3\right]\chi_i\epsilon^{\nu}_{\lambda}(q)q'_{\nu}\bar{u}_{s'}(p')\gamma_5\gamma^{\mu}u_s(p)\frac{i\left(p'_{\mu}-p_{\mu}\right)}{{\left(p'-p\right)}^2-m_{\pi}^2+i\epsilon}\nonumber \\
    &+\frac{ieg_A}{4F_{\pi}}\chi_f^{\dagger}\left[\tau_a,\tau_3\right]\chi_i\epsilon^{\nu}_{\lambda}(q)\bar{u}_{s'}(p')\gamma_5\gamma^{\mu}u_s(p)\frac{i\left(p'_{\nu}-p_{\nu}\right)\left(p'_{\mu}-p_{\mu}\right)}{{\left(p'-p\right)}^2-m_{\pi}^2+i\epsilon}\ ,
\end{align}
\begin{align}
    i\mathcal{M}^{(2)}=&\frac{-eg_A}{2F_{\pi}}\chi_f^{\dagger}\tau_a\bigg[\frac{c_6}{2}\left(q_{\nu}\epsilon_{\mu,\lambda}(q)-q_{\mu}\epsilon_{\nu,\lambda}(q)\right)\tau_3\nonumber \\
    &+\frac{c_7}{4}\left(q_{\nu}\epsilon_{\mu,\lambda}(q)-q_{\mu}\epsilon_{\nu,\lambda}(q)\right)\bigg]\chi_i{q'}^{\rho}\bar{u}_{s'}(p')\gamma_5\gamma_{\rho}\frac{i}{\left(\slashed{q}+\slashed{p}\right)-m_N+i\epsilon}\sigma^{\mu\nu}u_s(p)\nonumber \\
    &+\frac{-eg_A}{2F_{\pi}}\chi_f^{\dagger}\bigg[\frac{c_6}{2}\left(q_{\nu}\epsilon_{\mu,\lambda}(q)-q_{\mu}\epsilon_{\nu,\lambda}(q)\right)\tau_3\\
    &+\frac{c_7}{4}\left(q_{\nu}\epsilon_{\mu,\lambda}(q)-q_{\mu}\epsilon_{\nu,\lambda}(q)\right)\bigg]\tau_a\chi_i{q'}^{\rho}\bar{u}_{s'}(p')\sigma^{\mu\nu}\frac{i}{\left(\slashed{p}-\slashed{q}'\right)-m_N+i\epsilon}\gamma_5\gamma_{\rho}u_s(p)\ ,\nonumber
\end{align}
where superscript stands for chiral order. Now those invariant scalar functions can be extracted from the above amplitudes:
\begin{eqnarray}
    \mathcal{A}_1^{+}&=&-\frac{ieg_A{m}_N}{2F_{\pi}}\left(\frac{1}{u-m_N^2}+\frac{1}{s-m_N^2}\right)-\frac{ieg_A{c}_6}{F}\left(\frac{2m_N^2}{u-m_N^2}+\frac{2m_N^2}{s-m_N^2}+1\right)\ ,\nonumber \\
    \mathcal{A}_1^{0}&=&-\frac{ieg_A{m}_N}{2F_{\pi}}\left(\frac{1}{u-m_N^2}+\frac{1}{s-m_N^2}\right)-\frac{ieg_A{c}_7}{2F}\left(\frac{2m_N^2}{u-m_N^2}+\frac{2m_N^2}{s-m_N^2}+1\right)\ ,\nonumber \\
    \mathcal{A}_1^{-}&=&-\frac{ieg_A{m}_N}{2F_{\pi}}\left(-\frac{1}{u-m_N^2}+\frac{1}{s-m_N^2}\right)-\frac{ieg_A{c}_6}{F}\left(-\frac{2m_N^2}{u-m_N^2}+\frac{2m_N^2}{s-m_N^2}\right)\ ,\nonumber \\
    \mathcal{A}_2^{+}&=&\frac{ieg_A{m}_N}{4F_{\pi}P\cdot q}\left(\frac{1}{u-m_N^2}-\frac{1}{s-m_N^2}\right)\ ,\label{bpes}\\
    \mathcal{A}_2^{0}&=&\frac{ieg_A{m}_N}{4F_{\pi}P\cdot q}\left(\frac{1}{u-m_N^2}-\frac{1}{s-m_N^2}\right)\ ,\nonumber \\
    \mathcal{A}_2^{-}&=&-\frac{ieg_A{m}_N}{4F_{\pi}P\cdot q}\left(\frac{1}{u-m_N^2}+\frac{1}{s-m_N^2}+\frac{4}{t-m_{\pi}^2}\right)\ ,\nonumber \\
    \mathcal{A}_3^{+}&=&\frac{eg_A{c}_6m_N}{F_{\pi}}\left(\frac{1}{u-m_N^2}-\frac{1}{s-m_N^2}\right)\ ,\nonumber \\
    \mathcal{A}_3^{0}&=&\frac{eg_A{c}_7m_N}{2F_{\pi}}\left(\frac{1}{u-m_N^2}-\frac{1}{s-m_N^2}\right)\ ,\nonumber \\
    \mathcal{A}_3^{-}&=&\frac{eg_A{c}_6m_N}{F_{\pi}}\left(-\frac{1}{u-m_N^2}-\frac{1}{s-m_N^2}\right)\ ,\nonumber \\
    \mathcal{A}_4^{+}&=&-\frac{eg_A{c}_6m_N}{F_{\pi}}\left(\frac{1}{u-m_N^2}+\frac{1}{s-m_N^2}\right)\ ,\nonumber \\
    \mathcal{A}_4^{0}&=&-\frac{eg_A{c}_7m_N}{2F_{\pi}}\left(\frac{1}{u-m_N^2}+\frac{1}{s-m_N^2}\right)\ ,\nonumber \\
    \mathcal{A}_4^{-}&=&-\frac{eg_A{c}_6m_N}{F_{\pi}}\left(-\frac{1}{u-m_N^2}+\frac{1}{s-m_N^2}\right)\ \ .
\end{eqnarray}

\subsection{Partial wave projection}

It is convenient to perform partial wave projection using the helicity formalism proposed in Ref.~\cite{Jacob2000a}. 
To that end, one can substitute the photon polarization vector \(\epsilon_\mu(q)\), the nucleon spinors \(u(p)\) and \(\bar{u}(p^\prime)\) in Eq.~\eqref{eq:LorAmp} by their helicity eigenstates \(\epsilon_\mu(q,\lambda_2)\), \(u(p,\lambda_1)\) and \(\bar{u}(p^\prime,\lambda_3)\) in the center of mass frame
\footnote{The spinor satisfies \(\vec{\Sigma}\cdot\frac{\vec{p}}{|\vec{p}|}u_{\pm}(p)=\pm u_{\pm}(p)\), and polarization vector satisfies \(\epsilon_{\pm}(q)=\frac{1}{\sqrt{2}}\left(\epsilon_1(q)\pm i\epsilon_2(q)\right)\).},
where \(\lambda_i\) (\(i=1,2,3\)) stand for helicity quantum numbers of initial nucleon, photon and final nucleon, in order. For each set of  helicity quantum numbers, denoted by \(H_s\equiv \{\lambda_1\lambda_2\lambda_3\} \), there is a helicity amplitude \(\mathcal{M}^I_{H_s}\), which can be expanded as
\footnote{It is worth stressing that there are in total 8 helicity amplitudes, nevertheless, only 4 of them are independent thanks to symmetries under parity and time reversion transformation.}
\begin{eqnarray}
    \mathcal{M}_{H_s}^I(s,t)=16\pi\sum_{J=M}^\infty(2J+1)\mathcal{M}_{H_s}^{IJ}(s)\,{d}_{\lambda\lambda^\prime}^J(\theta)\ ,
\end{eqnarray}
where \(M=\lambda \), \(\lambda\equiv\lambda_1-\lambda_2\) and \(\lambda^\prime\equiv\lambda_3\).
\(d^J(\theta)\) is the standard Wigner \(d\)-function. 
By imposing the orthonormal properties of the \(d^J\) functions, the partial wave helicity amplitudes \(\mathcal{M}_{H_s}^{IJ}(s)\) in the above equation may be projected, i.e.
\begin{align}\label{pwamp}
    \mathcal{M}^{IJ}_{H_s}(s)=\frac{1}{32\pi}\int_{-1}^{1} d\cos\theta \mathcal{M}^{I}_{H_s}(s,t)d_{\lambda,\lambda'}^{J}(\theta)\ .
\end{align}

The partial wave amplitude with \(I=\frac{1}{2}\), \(J=\frac{1}{2}\) and \(L=0\) (denoted by \(S_{11}\) in \(L_{2I2J}\) convention) is obtained via
\begin{eqnarray}
    \mathcal{M}(S_{11})=\bigg(\mathcal{M}^{I=\frac{1}{2}J=\frac{1}{2}}_{+++}+\mathcal{M}^{I=\frac{1}{2}J=\frac{1}{2}}_{++-}\bigg)\ ,
\end{eqnarray}
which carry certain parity
\footnote{The positive direction particle is the direction of nucleon and the negative direction state is defined through \(\lvert-p_z,\lambda\rangle=e^{-i\pi J_z}e^{-i\pi J_y}\lvert p_z,\lambda\rangle \), which differs in a phase with the case in Ref.~\cite{Jacob2000a}.}
and the helicity indices \(\lambda_i=\pm\frac{1}{2}\) or \(\pm 1\) are abbreviated by \(\pm \).

\subsection{Singularities of partial wave amplitudes}
\subsubsection{The analytic structure of partial wave amplitudes}

To illustrate the analytic structure of the partial wave amplitudes, we rewrite the partial wave projection formula in Eq.~\eqref{pwamp} in the following form
\begin{eqnarray}\label{pwaef}
    \mathcal{M}^{IJ}_{H_s}(s)=\frac{1}{32\pi}\int_{t_{\min}}^{t_{\max}} \sum_{i=1}^4\big[{(\mathcal{G}_{H_s}^J)}_i\mathcal{A}_i^I(s,t)\big]{\rm d}t\ ,
\end{eqnarray}
where the invariant amplitude \(\mathcal{M}^I_{H_s}\) has replaced by its Lorentz-decomposed expression given in Eq.~\eqref{eq:LorAmp} and \(t_{\min},\ t_{\max}\) correspond to \(\cos{\theta}=\pm 1\) through Eq.~\ref{eq_rhot}.
Furthermore, the scalar functions \({(\mathcal{G}_{H_s}^{J})}_i\) (\(i=1,\cdots 4\)) are defined by
\begin{eqnarray}\label{eq:scalarH}
    {(\mathcal{G}_{H_s}^{J})}_i\equiv\bar{u}L^{i}_{\mu}u\epsilon^\mu\frac{d^J_{\lambda_1\lambda^\prime}(s,t)}{s\rho_{\pi N}\rho_{\gamma N}}\ ,
\end{eqnarray}
where \(L^{i}_{\mu}\) can be found in Eq.~\eqref{ltd}. In what follows, we proceed to discuss the analytic structure with the help of Eq.~\eqref{pwaef}. Note here that the Mandelstam variable \(t\) is related to the cosine of the scattering angle \(\theta \) via
 \begin{equation}\label{eq_rhot}
    t=2m_N^2-\frac{\left(s+m_N^2\right)\left(s+m_N^2-m_{\pi}^2\right)}{2s}+s\rho_{\pi N}\rho_{\gamma N}\frac{\cos\theta}{2}\ .
\end{equation}

On the one hand, It should be emphasized that the functions \({(\mathcal{G}_{H_s}^{J})}_{i=\ldots ts,4}\) rely merely on the kinematical structures of the scattering amplitudes, regardless of the dynamics of the system under consideration.
Therefore, they are model-independent and can be calculated straightforwardly for any partial wave quantum numbers of \(J\).
In Appendix~\ref{PWAS2}, for \(J={1}/{2}\) and \(H_s\)=\{++--,+++\}, all the explicit expressions of \({(\mathcal{G}_{H_s}^{J})}_{i}\) are listed for the sake of easy reference.
It can be observed that \({(\mathcal{G}_{H_s}^{J=\frac{1}{2}})}_{i}\) in \(S_{11}\) channel are just polynomials of \(t\).

On the other hand, information on the dynamics are completely encoded in the scalar amplitudes \(\mathcal{A}^I_{i}(s,t)\).
In our tree-level ChPT calculation, they are represented by the results shown in Subsection~\ref{lddabcd}, which are composed of contact terms, \(t\)-channel pion-pole, \(s\)- and \(u\)-channel nucleon-exchange contributions.
The contact term and \(s\)-channel nucleon exchange term are polynomials of \(t\), while \(t\)- and \(u\)-channel pole terms
\footnote{\(\frac{1}{P\cdot q}\) due to kinematical decomposition can be transformed into this kind of form.}
can be unified to a single type, \(1/(t-c)\), with \(c\) a function of \(s\).

Restricted to our tree-level calculation and with the above discussions, one can conclude there exist only one master integral:
\begin{eqnarray}\label{eq:int}
    \int_{t_{\min}}^{t_{\max}}\frac{t}{t-c}{\rm d}t=t_{\max}-t_{\min}+c\left[{\ln}\left(t_{\max}-c\right)-{\ln}\left(t_{\min}-c\right)\right]\ .
\end{eqnarray}
All other integrals are either trivial in the sense that they are integrations over polynomials of \(t\), being able to be reduced to the above integral by making use of the identity \(\frac{t^n}{t-c}=t^{n-1}+\frac{ct^{n-1}}{t-c}\) with \(n\) a positive integer.
In our current case, the constant \(c\) has three options, i.e.,
\(c\in \{m_\pi^2, s-m_N^2-m_\pi^2, 2s-2m_N^2-m_\pi^2\} \),
which result in three typical logarithms \(\mathcal{D}_i(s)\) after applying Eq.~\eqref{eq:int}.
We refer the readers to Appendix~\ref{PWAS2} for their explicit expressions. For the \(\mathcal{D}_i(s)\) except \(\mathcal{D}_3(s)\), which comes from kinematical decomposition, it should be mentioned that those logarithms are stemmed from the dynamics term \(1/(t-c)\), while their composite arguments could be square root functions originated from the kinematical limits of the integrations.
The logarithms and square root functions give rise to the partial-wave singularities to be discussed in the following subsections.

\subsubsection{Dynamical singularities}
\begin{figure}[H]\centering
    \begin{tikzpicture}[global scale = 0.55]
        \draw[ultra thick, orange](0,0) circle (4);
        \fill [white] (3.5,-3)to(4.5,-3)to(4.5,3)to(3.5,3)to(3.5,-3);
        \draw[dashed, very thick](3.5,-3)to(3.5,3);
         \node[right] at (3.5,-2){\LARGE\(s_2^-\)};
         \node[right] at (3.5,2){\LARGE\(s_2^+\)};
        \draw[->](-8.5,0)to(8.5,0);
        \node[above] at (8.5-0.5,0){\LARGE\(\infty\)};
        \draw[->](0,-5)to(0,5); \node[above] at (-8.5+0.5,0){\LARGE\(-\infty\)};
        \draw(4,0) circle(0.1);
        \fill[black] (4,0) circle(0.1);
        \draw[dashed, red] (4,0) circle(0.3);
        \node[below] at (4,-0.5) {\LARGE\(s_N\)};
        \draw[ultra thick, blue](2.5,0)to (-8,0);
        \node[above] at (2.5,0){\LARGE\(s_1\)};
        \draw[ultra thick, blue](4.5,0)to (8,0);
        \node[above] at (4.7,0){\LARGE\(s_R\)};
    \end{tikzpicture}
    \caption{Dynamical Singularities. \(s_N=m_N^2\), \(s_1=\frac{m_N}{m_{\pi}+m_N}\left(m_N^2-m_N{m}_{\pi}-m_{\pi}^2\right)\). }\label{dynsing}
\end{figure}
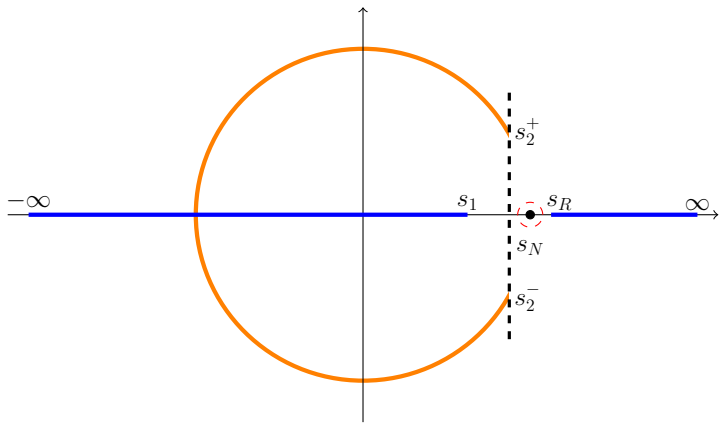

The generic dynamical singularities of the partial-wave photoproduction amplitude have been discussed in detail in Ref.~\cite{Kennedy1962}.
All possible singularities are displayed in Fig.~\ref{dynsing} and are briefly illustrated in the following.

\begin{itemize}
\item {unitarity cut:} \(s\in[s_R,\infty)\) on account of the \(s\)-channel continuous spectrum.
\item \(u\)-channel crossed cut:  \(s\in(-\infty,s_1]\) with \(s_1=\frac{m_N}{m_{\pi}+m_N}\left(m_N^2-m_N{m}_{\pi}-m_{\pi}^2\right)\) due to the \(u\)-channel continuous spectrum for \({u\geq(m_N+m_\pi)}^2\).
\item \(t\)-channel crossed cut: I. The arc, with branch points located at \(s_2^\pm=m_N^2-\frac{3}{2}m_{\pi}^2\pm\frac{i}{2} m_{\pi } \sqrt{\frac{44}{5}m_N^2-9 m_{\pi }^2}\), stems from \(t\)-channel continuous spectrum for \(4m_{\pi}^2\leq t \leq 4m_N^2\);
\footnote{This arc is not a circle arc. See Ref.~\cite{Kennedy1962} for detailed discussion.} 
II\@. The \(t\)-channel continuous spectrum above \(4m_N^2\) yields the cut \(s\in(-\infty,0]\). 
\item Trivial cut: {\(s\in(-\infty,0]\) generated by the logarithms.}
\item Discrete term: located at \(s=m_N^2\equiv s_N\) and induced by the \(t\)-channel single pion exchange as well as the \(u\)-channel single nucleon exchange.
\footnote{Actually, this isolated branch point singularity disappears after appropriately arranging the logarithms in the partial wave amplitudes. However, a singularity at \(m_N^2\) will still be there due to kinematical properties, which will be discussed in the next subsection.}.
\end{itemize}

Let us come back to our special case under consideration.
Since the continuous spectrums are absent for a tree-level calculation, we meet only with the dynamical singularities of the trivial cut and the discrete term.

\subsubsection{Kinematical singularities}
Aside from the above-mentioned dynamical singularities, there exist additional kinematical singularities for an inelastic scattering process with spinors.
The kinematical singularities are caused by the square-root and/or logarithmic functions appearing in the partial wave amplitudes.
Kinematical cuts are introduced when the arguments of those two kinds of functions are negative.
All the involved arguments together with their corresponding negative domains are listed in Table~\ref{sinre}. 
    \begin{table}[H]\small
        \centering
        \caption{Arguments causing singularities}\label{sinre}
        \begin{threeparttable}
%            \item[]    
            \begin{tabular}{@{}ll}
                \toprule
                Arguments&Negative Domain\\
                \midrule
                \(s-s_R\)&\(\left(-\infty,s_R\right)\) \\
                \midrule
                \(s-s_L\)&\(\left(-\infty,s_L\right)\) \\
                \midrule
                \(s\)&\(\left(-\infty,0\right)\) \\
                \midrule
                \(s+m_N^2-m_{\pi}^2-\sqrt{s-s_R}\sqrt{s-s_L}\)&- \\
                \midrule
                \(s+m_N^2-m_{\pi}^2+\sqrt{s-s_R}\sqrt{s-s_L}\)&\(\left(-\infty,0\right)\) \\
                \midrule
                \(3s+m_N^2-m_{\pi}^2-\sqrt{s-s_R}\sqrt{s-s_L}\)&\(\left(-\infty,\frac{1}{2}\left(m_{\pi}^2-2m_N^2\right)\right)\) \\
                \midrule
                \(3s+m_N^2-m_{\pi}^2+\sqrt{s-s_R}\sqrt{s-s_L}\)&\(\left(-\infty,0\right)\) \\
                \midrule
                \(s-m_N^2+m_{\pi}^2-\sqrt{s-s_R}\sqrt{s-s_L}\)&\(\left(0,s_L\right)\) \\
                \midrule
                \(s-m_N^2+m_{\pi}^2+\sqrt{s-s_R}\sqrt{s-s_L}\)&\(\left(-\infty,s_L\right)\) \\
                \bottomrule
            \end{tabular}
        \end{threeparttable}
    \end{table}
It should be pointed out that how these functions are organized in the way that does not affect the value in the physical region but may affect the values in complex plane.
Here we give an example to illustrate it:
\begin{description}
    \item[\(\sqrt{\left(s-s_R\right)\left(s-s_L\right)}\) Case:] There are two cuts. One goes from \(s_L\) to \(s_R\) and the other is an infinitely-long line, which is perpendicular to the real axis and passes the midpoint of \(s_L\) and \(s_R\).
    \item[\(\sqrt{s-s_R}\sqrt{s-s_L}\) Case:]There is just one cut stretching from \(s_L\) to \(s_R\) with the cuts below \(s_L\) cancelling each other. 
\end{description}
Meanwhile the values in the  physical region in the above two cases are the same.
In practice, we choose to expand the root functions in terms of power series and then continue them to the full complex plane.
In this way, all the kinematical singularities represent themselves as cuts lying on real axis.
And the logarithm functions in the form of \(\ln\frac{a}{b}\), whose arguments contain root functions, are recast to \({\ln}a-{\ln}b\) in order to avoid the circular cut in the complex plane.

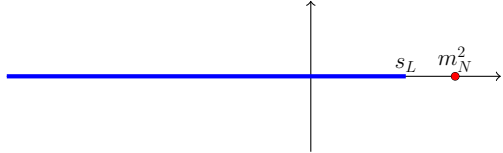
\begin{figure}[H]\centering
    \begin{tikzpicture}[global scale = 0.5]
        \draw[->](-8,0)to(5,0);
        \draw[->](0,-2)to(0,2);
        \draw[ultra thick, blue](-8,0)to (2.5,0);
        \node[above] at (2.5,0){\LARGE\(s_L\)};
        \draw (3.8,0) circle (0.1);
        \fill[red](3.8,0) circle (0.1);
        \node[above] at (3.8,0){\LARGE\(m_N^2\)};
    \end{tikzpicture}
    \caption{Kinematical Singularities}\label{kinsing}
\end{figure}

For the \(S_{11}\) channel, the cut between \(s_L\) and \(s_R\) disappears since \(\mathcal{M}^{I=\frac{1}{2}J=\frac{1}{2}}_{+++}\) and \(\mathcal{M}^{I=\frac{1}{2}J=\frac{1}{2}}_{++-}\) are conjugated with each other in this interval and this is not hard to be understood in observation of the explicit form of \({(\mathcal{G}_{H_s}^{J})}_i\) in Appendix~\ref{PWAS2}.
In addition, there is a pole-like singularity at \(m_N^2\) coming from the fact that \(\lim_{s\to m_N^2}\frac{\mathcal{D}_i}{\rho_{\gamma-N}}\), where the source of \(\frac{1}{\rho}\) can be seen in Eq. (\ref{pwaef}), diverge, meanwhile the limit of \(\lim_{s\to s_R}\frac{\mathcal{D}_i}{\rho_{\gamma-N}}\) is finite.
But this pole-like singularity appearing in the amplitude can be viewed as the branch point of the left hand cut starting from \(m_N^2\) in \(\mathcal{S}\) matrix since \(S_{\gamma\pi}=\sqrt{\rho_{\gamma N}\rho_{\pi N}}T\), where \(S_{\gamma\pi}\) is pion photoproduction \(\mathcal{S}\) matrix, or the branch point of electromagnetic unitarity cut of amplitudes.
The results of additional singularity in \(S_{11}\) channel are displayed in Figure~\ref{kinsing}.
As the result of kinematical singularities, we should include \(s\)-channel and contact diagrams besides \(t\)- and \(u\)-channel resonance exchanges in the estimation of \(\mathcal{M}_L\) at tree level.

\section{Numerical results and discussions}
We are now in the position to compare the dispersive representation of photoproduction amplitude given in Eq.~\eqref{eq:disRep} with experimental multipole amplitude data from Ref.~\cite{Workman2012} in \(S_{11}\) channel.
Based on our fitting results, the couplings of \(N^\ast(890)\) to \(\gamma N\) and \(\pi N\) can be extracted.

\subsection{The fitting procedure}

In our fitting procedure, there are three different kinds of parameters in Eq.~\eqref{eq:disRep}: the LECs involved in determination of \(\mathcal{M}_L(s)\), the subtraction constants in the auxiliary function \(\mathcal{D}(s)\) and the ones in the overall subtraction polynomial \(\mathcal{P}(s)\).
Firstly, the parameters in the Lagrangian appearing in \(\mathcal{M}_L(s)\) are chosen to be \(m_N=0.9383~{\rm MeV}\), \(m_{\pi}=0.1396~{\rm MeV}\), \(e=0.303\), \(g_A=1.267\), \(F_{\pi}=92.4~{\rm MeV}\), \(c_6={3.706}/{(4m_N)}\) and \(c_7={-0.12}/{(2m_N)}\)\footnote{Neglecting ChPT correction beyond tree level, the two LECs \(c_6\) and \(c_7\) can be related to the anomalous magnetic moments of the nucleon via
\begin{eqnarray}
  c_6=\frac{k_p+k_n}{2m_N}\quad ,
  c_7=\frac{k_p-k_n}{4m_N}\quad ,
\end{eqnarray}
with \(k_p\) and \(k_n\) being anomalous magnetic moments of proton and neutron, respectively. Since $k_p$ and $k_n$ are precisely determined by experiments~\cite{PDG2018}, one can infer the uncertainties of $c_6$ and $c_7$ must be negligible and shall hardly change our results.}.
Secondly, we set \(\tilde{\mathcal{P}}(s)=1\) and compute \(\mathcal{D}(s)\) by using the \(S_{11}\)-wave phase shift extracted from the \(\pi N\) \(S\) matrix given in Ref.~\cite{Wang2018}.
Two solutions of the \(\pi N\) \(S\) matrix are adopted: one corresponding to \(s_c=-1~\rm{GeV}^2\) and the other to \(s_c=-9~\rm{GeV}^2\) with \(s_c\) being a cut off parameter therein.
Note that it should be a good approximation for a single-channel case that the integrations in Eqs. (\ref{eq:omnes}) and (\ref{eq:disRep}) are performed up to \(2.095\rm GeV^2\) rather than to infinity.
Lastly, the constants in the \(\mathcal{P}\) are left as fitting parameters.
\footnote{\(\tilde{\mathcal{P}}\) can always be chosen to be \(1\) in Eq.~\eqref{eq:rhue}.}
Here we only consider two fit cases: Fit I with \(\mathcal{P}(s)=a\) and Fit II \(\mathcal{P}(s)=a+b\, s\) while the subtraction points are set to be zero.

\begin{figure}[H]\centering
    \subfigure[Real Part]{
        \includegraphics[width=5in]{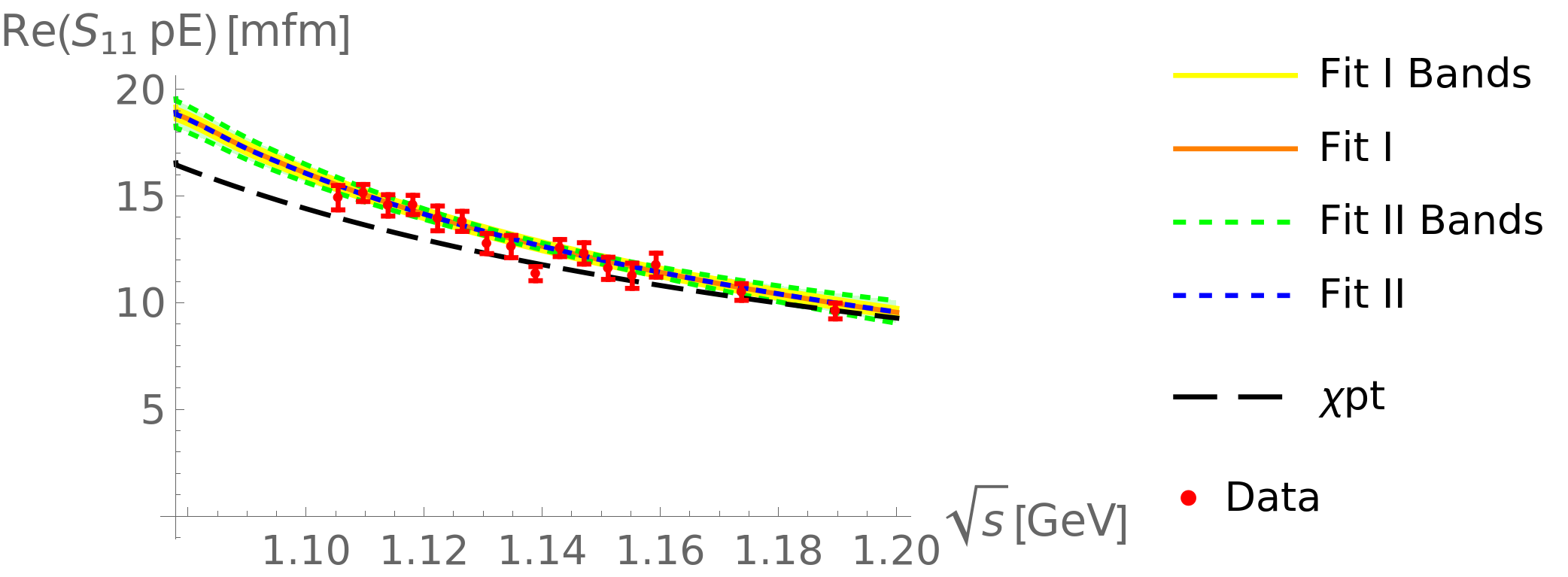}
    }
    \subfigure[Imaginary Part]{
        \includegraphics[width=5in]{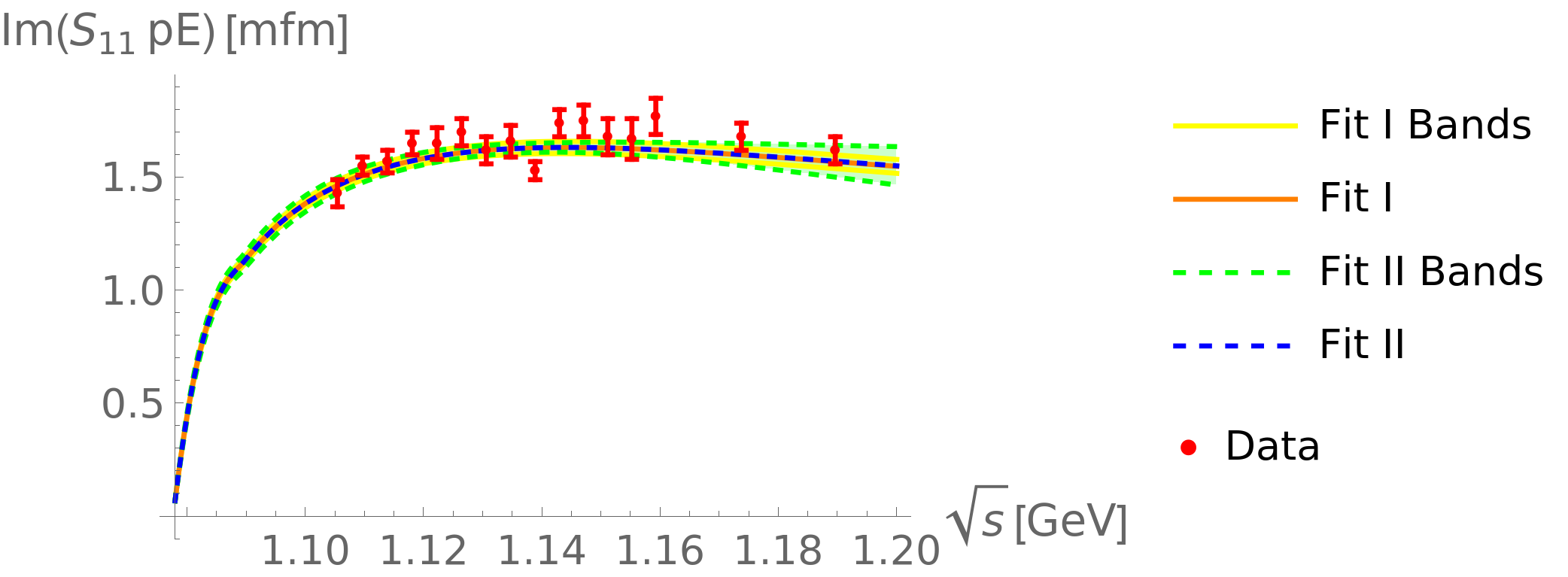}
    }
    \caption{\(p\) Target. Upper panel: real part of the \(S_{11}\) electric multipole; Lower panel: imaginary part of the \(S_{11}\) electric multipole. The solid orange and dashed blue lines represent our dispersive descriptions based on Fit I and Fit II, respectively. Yellow solid line and green dashed line represent the error bands of Fit I and Fit II. For comparison, the chiral result of the real part of the multipole is also shown, corresponding to the black long dashed line.}\label{ufrp}
\end{figure}

\begin{figure}[H]\centering
    \subfigure[Real Part]{
        \includegraphics[width=5in]{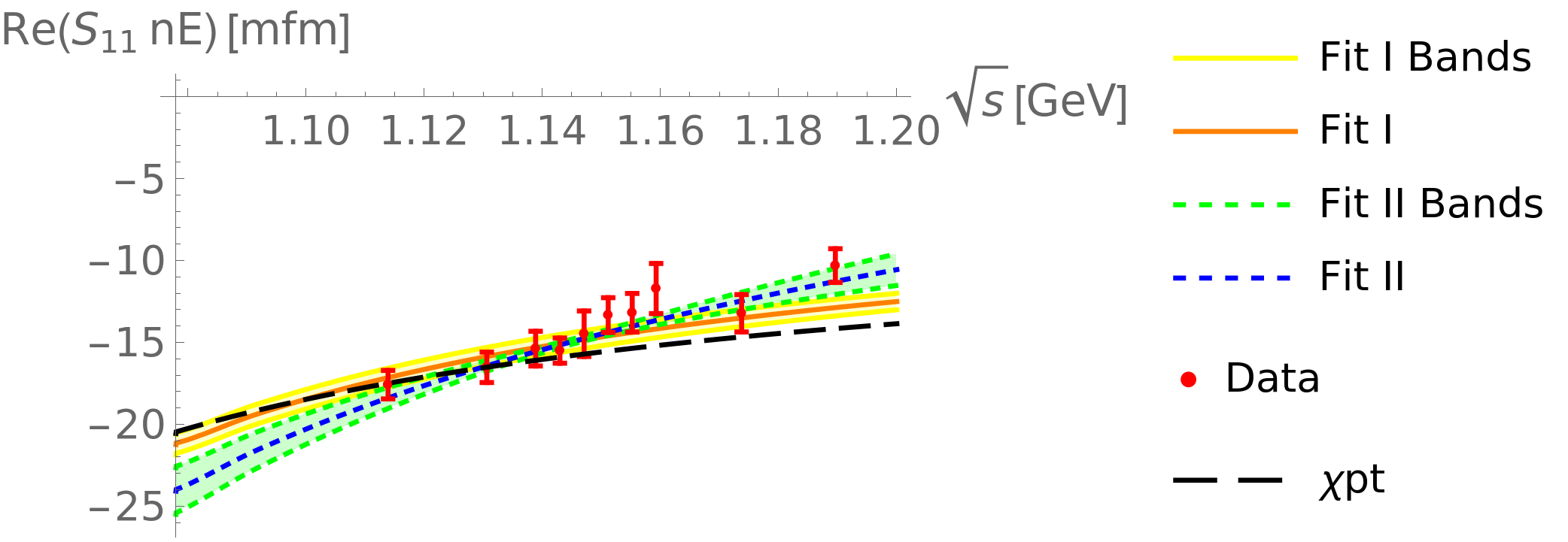}
    }
    \subfigure[Imaginary Part]{
        \includegraphics[width=5in]{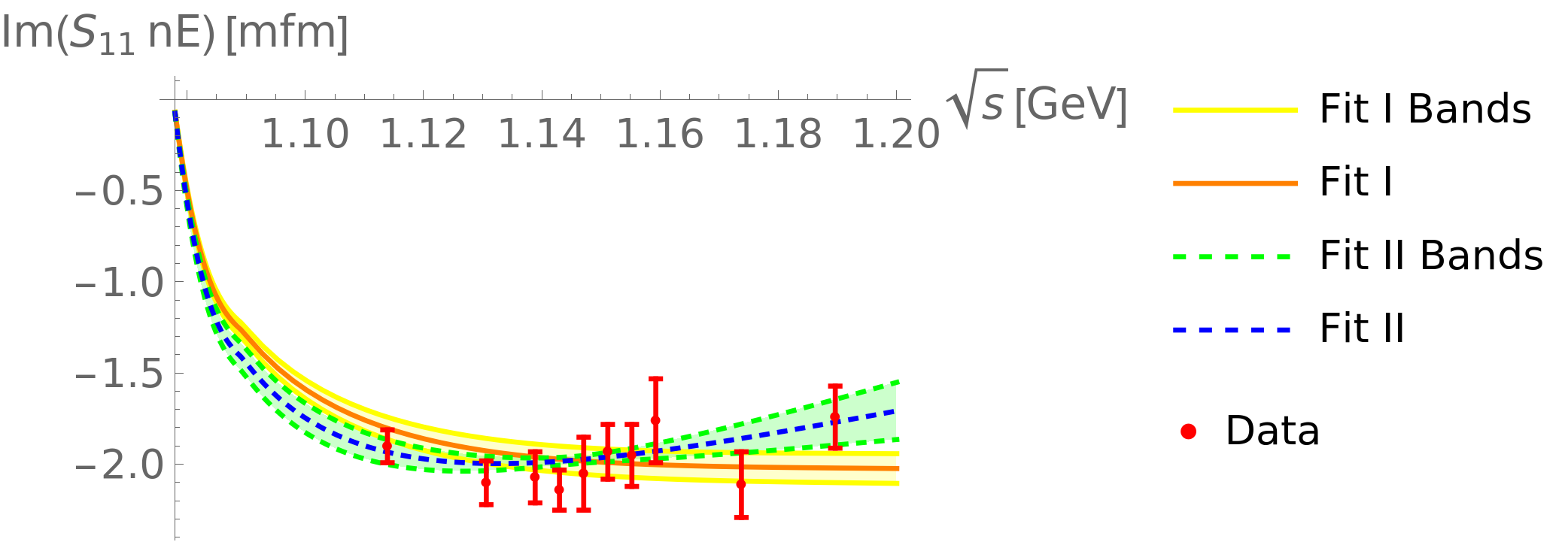}
    }
\caption{\(n\) Target. Same definitions as in Fig.~\ref{ufrp}}\label{ufrn}
\end{figure}

We perform a fit to the data points on the multipole amplitudes
\footnote{The relation between multipole amplitudes and our amplitudes can be established through traditional CGLN convention, which can be found in Appendix~\ref{ap:CGLN}.},
which are traditionally denoted by \(S_{11}\) with suffixes of target type (n or p) and electromagnetic transition (\(E\):electric, \(M\):magnetic), from \(\pi N\) threshold to \(1.440~\rm GeV^2\) just below the peak of \(\Delta(1232)\).
The fit results for both proton (p) and neutron (n) targets are displayed in Figures~\ref{ufrp} and~\ref{ufrn}, respectively. 
For comparison, in Fig.~\ref{ufrp} and Fig.~\ref{ufrn}, we also show the \(\mathcal{O}(q^2)\)  chiral results of the real parts of multipole amplitudes. 
As expected, the chiral results only describe the data very well at low energies close to threshold. 
The values of the fit parameters are collected in Table~\ref{pfrq}.

\begin{table}[H]\footnotesize
    \centering
    \caption{Results of the fit parameters. \(a\) is dimensionless and the unit of \(b\) is \(\rm{GeV}^{-1}\).}\label{pfrq}
    \begin{threeparttable}
        \begin{tabular}{@{}cccrc}
            \toprule
            Target&Case&Parameter&Value&\(\chi^2/d.o.f\) \\
            \midrule
            \multirow{3}*{p}&Fit I&\(10^2\times a\)&\(-0.0712\pm0.1334\)&\(1.58\) \\
            \cline{2-5}
            &\multirow{2}*{Fit II}&\(10^2\times a\)&\(0.0287\pm3.2525\)&\multirow{2}*{\(1.63\)} \\
            \cline{3-4}
            &&\(10^2\times b\)&\(-0.210\pm2.504\)&\\
            \midrule
            \multirow{3}*{n}&Fit I&\(10^2\times a\)&\(-1.43\pm0.35\)&\(1.22\) \\
            \cline{2-5}
            &\multirow{2}*{Fit II}&\(10^2\times a\)&\(12.50\pm6.90\)&\multirow{2}*{\(0.643\)} \\
            \cline{3-4}
            &&\(10^2\times b\)&\(-10.4\pm5.2\)&\\
            \bottomrule
        \end{tabular}
    \end{threeparttable}
\end{table}

For Fit I, our results are in good agreement with the experimental data in the sense that the averaged \(\chi^2\) are close to one, \(\chi^2/{d.o.f}= 1.58\) for the \(p\) target and  \(\chi^2/{d.o.f}= 1.22\) for the \(n\) target. 
{As can be seen from Table~\ref{pfrq}, the modulus of the central value of $a$ is close to zero in the proton target case, while it is $1.43$ in the neutron target case. That is due to the fact that the electric multipoles calculated from BChPT in the proton case can already well describe the experimental data, enforcing a nearly-zero contribution from the subtraction polynomial in the fitting procedure and further resulting a nearly-zero value of $a$. However, in the neutron case the discrepancy between BChPT results and experimental data are larger compared to the proton case, which leads to a larger central value (modulus) of $a$ subsequently.}

Fit II is performed by using a rank-2 subtraction polynomial with two parameters $a$ and $b$.
Compared to Fit I, the qualities of Fit II are improved, which is under our expectation since one more free parameter is involved in the fit procedure. However, the fitting parameters \(a\) and \(b\) of Fit II are highly correlated with a correlation coefficient which is nearly \(-1\).
Thus, Fit I is more advisable. 

\subsection{Analytic continuation and extraction of the \texorpdfstring{\(N^\ast(890)\)}{Lg} couplings}
In the above subsection, all the involved parameters in the dispersive \(S\)-wave photoproduction amplitude \(\mathcal{M}(s)\) have been determined.
Since \(N^\ast(890)\), as a subthreshold resonance, is located on the second Riemann sheet (RS), one needs to perform analytic continuation in order to extract its couplings to the \(\gamma N\) and \(\pi N\) systems. 

The amplitude on the second RS can be deduced via
\begin{eqnarray}
    \mathcal{M}^{{\rm II}}(s)=\frac{\mathcal{M}(s)}{\mathcal{S}(s)}\ ,
\end{eqnarray}
where \(\mathcal{M}(s)\) is the partial-wave photoproduction amplitude given in Eq.~\eqref{eq:disRep} and \(\mathcal{S}(s)\) corresponds to the \(S\) matrix of \(\pi N\) scattering with same quantum numbers as \(\mathcal{M}(s)\). If there exists a second RS pole located at \(z_R\), the \(S\) matrix can be approximated by 
 \begin{equation}
    \mathcal{S}(s)\approx \mathcal{S}'(z_{{\rm R}})\left(s-z_{{\rm R}}\right)
\end{equation}
in the vicinity of \(z_{{\rm II}}\). Thus,
\begin{align}
    \mathcal{M}^{{\rm II}}(s)=\frac{\mathcal{M}(s)}{\mathcal{S}'(z_{{\rm R}})\left(s-z_{{\rm R}}\right)}\ .\label{aac}
    \end{align}
On the other hand, the couplings of this second RS pole to the \(\gamma N\) and \(\pi N\) systems are defined as the residue via
\begin{align}
    \mathcal{M}^{{\rm II}}(s\rightarrow z_R)=\frac{g_{\gamma}g_{\pi}}{s-z_R}\ ,
\end{align}
with \(g_\gamma \) and \(g_\pi \) denoting the \(\gamma N\) and \(\pi N\) couplings, respectively.
Compared to Eq.~\eqref{aac}, one obtains
\begin{align}\label{eq:gggp}
    g_{\gamma}g_{\pi}=&\frac{\mathcal{M}(z_{{\rm R}})}{\mathcal{S}'(z_{{\rm R}})}\ .
    \end{align}
The \(\pi N\) coupling can also be extracted from elastic \(\pi N\) scattering, i.e.,
\begin{align}\label{eq:gpi2}
    g_{\pi}^2=\frac{\mathcal{T}(z_{{\rm R}})}{\mathcal{S}'(z_{{\rm R}})}\ ,
\end{align}
where \(\mathcal{T}\) is the corresponding partial-wave \(\pi N\) scattering amplitude.

Now we proceed with the numerical calculation of the couplings of \(N^*(890)\).
According to Eqs.~\eqref{eq:gggp} and~\eqref{nbmp}, the pion photoproduction \(N^*(890)\) residue couplings, i.e. \(g_{\gamma}g_{\pi}\), can be extracted from multipole amplitudes.
In the meantime, \(g_{\pi}^2\) can be computed by using Eq.~\eqref{eq:gpi2}, which was already done in Ref.~\cite{Wang2018}.
Results of the couplings are listed in Table~~\ref{resresult1}.
The results based on Fit II are also shown to check the stability of the obtained values.
We employed two solutions of the pole position of the \(N^*(890)\), \(\sqrt{s}=0.882-0.190i\) corresponding to the cutoff \(s_c=-1\)~GeV and \(\sqrt{s}=0.960-0.192i\) to \(s_c=-9\)~GeV\@;
See Ref.~\cite{Wang2018} for detailed explanation.

\begin{widetext}
    \begin{table}[H]\footnotesize\centering
        \centering
        \caption{Results of \(g_{\gamma}g_{\pi}\) and \(g_{\pi}^2\). Pole position, moduli and phase are in \(\rm{GeV}\), \(10^{-2}\times \rm{GeV^2}\) and degrees, in order. \(g_{\pi}^2\) are the same for p target and n target due to the isospin symmetry.}\label{resresult1}
        \begin{threeparttable}
            \begin{tabular}{@{}crrrrrrr}
                \toprule
                &&\multicolumn{4}{c}{\(g_{\gamma}g_{\pi}\)}&\multicolumn{2}{c}{\multirow{2}{*}{\(g_{\pi}^2\)}}\\
                \cline{3-6}
                &&\multicolumn{2}{c}{Fit I}&\multicolumn{2}{c}{Fit II}&&\\
                \midrule
                Target&Pole Position&Moduli&Phase&Moduli&Phase&Moduli&Phase\\
                \midrule
                \multirow{2}*{p}&\(0.882-0.190i\)&\(\left(1.212\pm0.014\right)\)&\(-79.2\pm1.3\)&\(1.203\pm 0.302\)&\(-78.9\pm 11.4\)&\(19.7\pm 0.3\)&\(32.6\pm1.0\) \\
                \cline{2-8}
                &\(0.960-0.192i\)&\(\left(1.467\pm0.016\right)\)&\(-71.3\pm0.9\)&\(1.459\pm 0.279\)&\(-71.2\pm3.5\)&\(21.4\pm 0.2\)&\(33.6\pm 0.8\) \\
                \midrule
                \multirow{2}*{n}&\(0.882-0.190i\)&\(\left(0.6416\pm0.0265\right)\)&\(111\pm7\)&\(2.025\pm 0.731\)&\(81.4\pm 6.9\) \\
                \cline{2-6}
                &\(0.960-0.192i\)&\(\left(1.111\pm0.050\right)\)&\(103\pm3\)&\(2.342\pm 0.605\)&\(98.0\pm 1.5\) \\
                \bottomrule
            \end{tabular}
        \end{threeparttable}
    \end{table}
\end{widetext}
In the extraction of \(g_{\gamma}g_{\pi}\) and \(g_{\pi}^2\) of \(N^*(890)\), \(z_R\) is treated as the \(N^*(890)\) pole position in the \(s\) plane, \(\mathcal{M}(z_R)\) can be obtained from the dispersion relation in Eq.~\eqref{eq:disRep} once \(\mathcal{P}\) is determined, \(\mathcal{T}(z_R)\) can be obtained through \(\mathcal{S}(z_R)=1+2i\rho_{\pi N}\mathcal{T}=0\) and \(\frac{1}{S'(z_R)}\) is just the residue of \(\mathcal{S}^{\rm II}\) from Ref.~\cite{Wang2018}.
But in order to compare the results of \(N^*(1535)\), which are extracted directly from multipole amplitudes parameterized in \(\sqrt{s}\) plane in Ref.\cite{Svarc2014}, the conventions should be consistent.
In \(S_{11}\) channel the following equation can be used to translate these residues from different conventions into residues directly extracted from multipole amplitudes in \(s\) plane.
\begin{equation}
    E^{I=\frac{1}{2}\ \rm II}_{0+}(s\rightarrow z_R)=-\sqrt{\frac{2}{3s}}\frac{g_{\gamma}g_{\pi}}{s-z_R}=-\sqrt{\frac{2}{3s}}\frac{g_{\gamma}g_{\pi}}{2\sqrt{z_R}(\sqrt{s}-\sqrt{z_R})}\ ,
\end{equation}
where \(R\) stands for \(N^*(890)\) or \(N^*(1535)\). In \(\rm S_{11}pE\) the moduli of residue is \(2.41\rm{mfm\cdot GeV^2}\) with phase \(120\degree \), meanwhile the magnitude of \(N^*(1535)\) residue coupling from Ref.~\cite{Svarc2014} is about \(0.736 \rm{mfm\cdot GeV^2}\) and phase is \(-27\degree \).
One can see the magnitude of the \(N^*(890)\) residue is larger than that of the \(N^*(1535)\) residue.
The \(|g_{\pi}^2|\) of \(N^*(890)\) is \(0.2\rm{GeV}^2\), and the one of \(N^*(1535)\), which is obtained by the value in Ref.~\cite{Arndt2006}, is \(0.08\rm{GeV}^2\).
The \(g_{\pi}^2\) of these two resonances may account for part of the reason why \(N^*(890)\) photoproduction residue is large, and using above results \(g_{\gamma}\) of these two resonances can be obtained.
The \(|g_{\gamma}|\) of \(N^*(890)\) is \(0.032\rm GeV\) meanwhile the one of \(N^*(1535)\) is \(0.024\rm GeV\) and one can see the magnitudes are almost the same.
One should notice that the results of n target are quiet unstable. 
The fact that data points are few and they have large error bars may account for the main reason.

We can also calculate the decay amplitudes \(\mathcal{A}^{\frac{1}{2}}\) at the \(N^\ast(890)\) pole position, which is related to the coupling \(g_{\gamma}\), using the formula given in Ref.~\cite{Svarc2014}:
\begin{equation}
    \mathcal{A}^{\frac{1}{2}}=g_{\gamma}\sqrt{\frac{\pi}{q_r^2m_N}\rho_{\gamma N}}\ ,
\end{equation}
where \(q_r\) is modulus of the photon momentum calculated at the resonance pole position.

Furthermore, we can obtain the partial widths of the \(N^*(890)\to\gamma N\) channel at the pole by following the next formula, which is from Ref.~\cite{Workman2013} and converted to our convention.
\begin{equation}
    \Gamma_{\gamma N}=\left\lvert\rho_{\gamma N}\frac{g_{\gamma}^2}{\sqrt{z_R}}\right\rvert \ ,
\end{equation}
where \(z_R\) is treated as \(N^*(890)\) pole position.
The values of the decay amplitudes \(\mathcal{A}^{\frac{1}{2}}\) and the partial decay width at the pole \(\Gamma_{\gamma N}\) are collected in Table~\ref{resresult2}.
\begin{widetext}
    \begin{table}[H]\footnotesize
        \centering
        \caption{Values of the decay amplitude (\(\mathcal{A}^{\frac{1}{2}}\)) and decay width (\(\Gamma_{\gamma N}\)) calculated at the \(N^\ast(890)\) pole position. Phase, \(\mathcal{A}^{\frac{1}{2}} \) and \(\Gamma_{\gamma N}\) are in degrees, \({\rm GeV^{-\frac{1}{2}}}\) and \(\rm MeV\), respectively.}\label{resresult2}
        \begin{threeparttable}
            \begin{tabular}{@{}crrrrrrr}
                \toprule
                \multirow{3}{*}{Target}&\multirow{3}{*}{Pole Position}&\multicolumn{4}{c}{\(\mathcal{A}^{\frac{1}{2}}\)}&\multicolumn{2}{c}{\(\Gamma_{\gamma N}\)}\\
                \cline{3-8}
                &&\multicolumn{2}{c}{Fit I}&\multicolumn{2}{c}{Fit II}&\multirow{2}{*}{Fit I}&\multirow{2}{*}{Fit II}\\
                \cline{3-6}
                &&Moduli&Phase&Moduli&Phase&&\\
                \midrule
                {p}&\(0.882-0.190i\)&\(0.165\pm 0.004\)&\(-129\pm2\)&\(0.165\pm0.043\)&\(-129\pm12\)&\(0.369\pm0.014\)&\(0.363\pm0.210\) \\
                \cline{2-8}
                &\(0.960-0.192i\)&\(0.191\pm0.004\)&\(-43.4\pm1.4\)&\(0.191\pm0.038\)&\(-43.3\pm3.9\)&\(0.396\pm0.013\)&\(0.391\pm0.168\) \\
                \midrule
                {n}&\(0.882-0.190i\)&\(0.0879\pm0.0043\)&\(61.7\pm8.2\)&\(0.277\pm0.102\)&\(31.4\pm7.4\)&\(0.103\pm0.011\)&\(1.03\pm0.89\) \\
                \cline{2-8}
                &\(0.960-0.192i\)&\(0.145\pm0.008\)&\(130\pm4\)&\(0.305\pm0.096\)&\(125\pm 3\)&\(0.227\pm0.023\)&\(1.01\pm0.73\) \\
                \bottomrule
            \end{tabular}
        \end{threeparttable}
    \end{table}
\end{widetext}
The \(|\mathcal{A}^{\frac{1}{2}}|\) of \(N^*(890)\) is larger than the one of \(N^*(1535)\), which is \(0.074 \rm GeV^{-\frac{1}{2}}\) with the phase being \(-17\degree \) in \(\rm S_{11}pE\) from Ref.~\cite{Svarc2014} but the decay widths at the pole are almost the same regardless of the instability of n target results.

\section{Summary}
In this paper, we have performed a careful dispersive analysis of the process of single pion photon production off the nucleon, in the \(S_{11}\) wave of the final pion-nucleon system.
In such a dispersive representation, the right-hand cut contribution can be related to an Omn\'es solution, which takes the elastic \(\pi N\) phase shifts as inputs, and hence is known up to a polynomial.
On the other hand, we estimate the left-hand cut contribution by making use of the \(\mathcal{O}(q^2)\) tree amplitudes taken from chiral perturbation theory.
A detailed discussion on how to establish a proper analytic structure of the partial-wave pion photon production amplitude is also presented for easy reference in future.
To pin down the free parameters in the dispersive amplitude, we perform fits  to experimental data of multipole amplitudes in the channels indicated by \(\rm S_{11}pE\) and \(\rm S_{11}nE\) for the energies ranging from \(\pi N\) threshold to \(1.440~\rm GeV^2\). 

It is found that the experimental data can be well described by the dispersive amplitude with only one free subtraction parameter.
We then continue the dispersive amplitude to the second Riemann sheet for the purpose of being able to extract the couplings of \(N^\ast \) to the \(\gamma N\) and \(\pi N\) systems, which are denoted by \(g_\gamma \) and \(g_\pi \), respectively.
Based on the obtained value of \(g_\gamma g_\pi \), the modulus of the corresponding residue of the multipole amplitude (\({\rm S_{11}pE}\)) at the \(N^\ast(890)\) pole position turns out to be \(2.41\rm{mfm\cdot GeV^2}\), which is much larger than the modulus of the residue of \(N^\ast(1535)\), i.e. \(0.736 \rm{mfm\cdot GeV^2}\)~\cite{Svarc2014}.
That means the strength of the interaction of \(N^\ast(890)\) with \(\pi N\) system is stronger, compared to the one regarding \(N^\ast(1535)\).
It’s physically reasonable and within expectation since \(N^\ast(890)\) is supposed to be composed of \(\pi N\) system and \(N^\ast(1535)\) has tiny coupling with \(\pi N\) as we all know.
The results provides further evidence of existence of \(N^\ast(890)\).
As byproducts, the decay amplitude and the decay width at the \(N^\ast(890)\) pole position \(\mathcal{A}_h\) and the \(\Gamma_{\gamma N}\) are obtained for future reference.

\section*{Acknowledgments}
This work is supported by National Nature Science Foundations of China (NSFC) under Contract number 11905258, 11975028 and 10925522, and by the Fundamental Research Funds for the Central Universities under No. 531118010379.
The authors are grateful to Y. F. Wang for valuable advises.

\begin{appendices}
 
\section{Partial wave amplitude}\label{PWAS2}
The functions \( {(\mathcal{G}_{H_s}^{J})}_i\) defined in Eq.~\eqref{eq:scalarH} are shown for \(S_{11}\) wave in the following:

\begin{align}
   {( \mathcal{G}_{+++}^{J=\frac{1}{2}})}_1=&ik_1 \sqrt{s} \left(k_l k_r \left(m_N^2-s\right)-m_N^2\left(m_{\pi }^2+2 s\right)+m_N^4+s \left(-m_{\pi }^2+s+2t\right)\right)\times\nonumber \\
    &{\left(\left(m_N^2-s\right) \left(\left(m_N-m_{\pi}\right){}^2-s\right) \left(\left(m_N+m_{\pi }\right){}^2-s\right)\right)}^{-1}\ ,\\
    {( \mathcal{G}_{++-}^{J=\frac{1}{2}})}_1=&-ik_2 \sqrt{s} \left(k_l k_r \left(s-m_N^2\right)-m_N^2\left(m_{\pi }^2+2 s\right)+m_N^4+s \left(-m_{\pi }^2+s+2t\right)\right)\times\nonumber \\
    &{\left(\left(m_N^2-s\right) \left(\left(m_N-m_{\pi}\right){}^2-s\right) \left(\left(m_N+m_{\pi }\right){}^2-s\right)\right)}^{-1}\ ,\\
    {( \mathcal{G}_{+++}^{J=\frac{1}{2}})}_2=&i s \left(k_1 \sqrt{s}-k_2 m_N\right) \left(m_N^2\left(m_{\pi }^4-t \left(m_{\pi }^2+2 s\right)\right)+t m_N^4+s t\left(-m_{\pi }^2+s+t\right)\right)\times\nonumber \\
    &{\left(\left(m_N^2-s\right){}^2\left(\left(m_N-m_{\pi }\right){}^2-s\right) \left(\left(m_N+m_{\pi}\right){}^2-s\right)\right)}^{-1}\ ,\\
    {( \mathcal{G}_{++-}^{J=\frac{1}{2}})}_2=&i s \left(k_1 m_N-k_2 \sqrt{s}\right) \left(m_N^2\left(m_{\pi }^4-t \left(m_{\pi }^2+2 s\right)\right)+t m_N^4+s t\left(-m_{\pi }^2+s+t\right)\right)\times\nonumber \\
    &{\left(\left(m_N^2-s\right){}^2\left(\left(m_N-m_{\pi }\right){}^2-s\right) \left(\left(m_N+m_{\pi}\right){}^2-s\right)\right)}^{-1}\ ,\\
    {( \mathcal{G}_{+++}^{J=\frac{1}{2}})}_3=&-k_2 \left(m_N^2 \left(k_l k_r-m_{\pi }^2-2 s\right)+s\left(-k_l k_r-m_{\pi }^2+s+2 t\right)+m_N^4\right)\times\nonumber \\
    &\Biggl[ k_l k_r\left(\frac{2 k_1 \sqrt{s} m_N \left(m_{\pi }^2-t\right)}{k_2}-m_N^2 \left(m_{\pi }^2+2 s\right)+m_N^4+s\left(m_{\pi }^2+s\right)\right)-\nonumber \\
    &\left(m_N^2-s\right)\left(\left(m_N-m_{\pi }\right){}^2-s\right) \left(\left(m_N+m_{\pi}\right){}^2-s\right)\Biggl ]\times\nonumber \\
    &{\left(4 \left(m_N^2-s\right){}^2\left(s-\left(m_N-m_{\pi }\right){}^2\right){}^{3/2}\left(s-\left(m_N+m_{\pi }\right){}^2\right){}^{3/2}\right)}^{-1}\ ,\\
    {( \mathcal{G}_{++-}^{J=\frac{1}{2}})}_3=&k_1 \left(-m_N^2 \left(k_l k_r+m_{\pi }^2+2s\right)+s \left(k_l k_r-m_{\pi }^2+s+2 t\right)+m_N^4\right)\times\nonumber \\
    &\Biggl[k_l k_r \left(\frac{2 \sqrt{s} m_N \left(m_{\pi }^2-t\right)k_2}{k_1}-m_N^2 \left(m_{\pi }^2+2 s\right)+m_N^4+s \left(m_{\pi}^2+s\right)\right)+\nonumber \\
    &\left(m_N^2-s\right) \left(\left(m_N-m_{\pi}\right){}^2-s\right) \left(\left(m_N+m_{\pi}\right){}^2-s\right)\Biggl]\times\nonumber \\
    &{\left(4 \left(m_N^2-s\right){}^2\left(s-\left(m_N-m_{\pi }\right){}^2\right){}^{3/2}\left(s-\left(m_N+m_{\pi }\right){}^2\right){}^{3/2}\right)}^{-1}\ ,\\
    {( \mathcal{G}_{+++}^{J=\frac{1}{2}})}_4=&\left(k_l k_r \left(m_N^2-s\right)-m_N^2 \left(m_{\pi }^2+2s\right)+m_N^4+s \left(-m_{\pi }^2+s+2 t\right)\right)\times\nonumber \\
    &\Biggl[k_2 k_l k_r\left(-2 \sqrt{s} m_N \left(-m_{\pi }^2+2 s+t\right)-m_N^2 \left(m_{\pi}^2-2 s\right)+m_N^4+s \left(m_{\pi }^2-3 s\right)\right)+\nonumber \\
    &4 k_1 \sqrt{s}k_l k_r m_N^3-k_2 \left(m_N^2-s\right) \left(\left(m_N-m_{\pi}\right){}^2-s\right) \left(\left(m_N+m_{\pi}\right){}^2-s\right)\Biggl]\times\nonumber \\
    &{\left(4 \left(m_N^2-s\right){}^2\left(s-\left(m_N-m_{\pi }\right){}^2\right){}^{3/2}\left(s-\left(m_N+m_{\pi }\right){}^2\right){}^{3/2}\right)}^{-1}\ ,\\
    {( \mathcal{G}_{++-}^{J=\frac{1}{2}})}_4=&\left(k_l k_r \left(m_N^2-s\right)+m_N^2 \left(m_{\pi }^2+2s\right)-m_N^4+s \left(m_{\pi }^2-s-2 t\right)\right)\times\nonumber \\ 
    &\Biggl[2 k_2 \sqrt{s}k_l k_r m_N \left(2 m_N^2+m_{\pi }^2-2 s-t\right)+k_1 \left(m_N^2-s\right)\times\nonumber \\
    &\left(k_l k_r \left(m_N^2-m_{\pi }^2+3 s\right)-2 m_N^2 \left(m_{\pi}^2+s\right)+m_N^4+\left(m_{\pi }^2-s\right){}^2\right)\Biggl]\times\nonumber \\
    &{\left(4\left(m_N^2-s\right){}^2 \left(s-\left(m_N-m_{\pi}\right){}^2\right){}^{3/2} \left(s-\left(m_N+m_{\pi}\right){}^2\right){}^{3/2}\right)}^{-1}
\end{align}
with
\begin{flalign}
    k_l=&\sqrt{s-s_L}\ ,\nonumber \\
    k_r=&\sqrt{s-s_R}\ ,\nonumber \\
    k_1=&\sqrt{s+m_N^2-m_{\pi}^2-\sqrt{s-s_R}\sqrt{s-s_L}}\ ,\\
    k_2=&\sqrt{s+m_N^2-m_{\pi}^2+\sqrt{s-s_R}\sqrt{s-s_L}}\ .\nonumber
\end{flalign}
The amplitudes \(A_{i}(s,t)\) up to \(\mathcal{O}(q^2)\) contain following terms:
\begin{itemize}
    \item \(t\) channel pion exchange: \(\frac{1}{t-m_{\pi}^2}\);
    \item \(u\) channel nucleon exchange: \(\frac{1}{u-m_N^2}=\frac{1}{m_N^2+m_{\pi}^2-s-t}\);
    \item Kinematical decomposition: \(\frac{1}{P\cdot q}=\frac{4}{t-2 m_N^2-m_{\pi }^2+2s}\).
\end{itemize}
They lead to logarithm terms:
\begin{align}\label{eq:Di}
    \mathcal{D}_1=&\ln \left(-\sqrt{s-s_L} \sqrt{s-s_R}+m_N^2-m_{\pi }^2+s\right)\nonumber \\
    &-\ln \left(\sqrt{s-s_L} \sqrt{s-s_R}+m_N^2-m_{\pi }^2+s\right)\ ,\\
    \mathcal{D}_2=&\ln \left(-\sqrt{s-s_L} \sqrt{s-s_R}-m_N^2+m_{\pi }^2+s\right)\nonumber \\
    &-\ln \left(\sqrt{s-s_L} \sqrt{s-s_R}-m_N^2+m_{\pi }^2+s\right)\ ,\\
    \mathcal{D}_3=&\ln \left(\sqrt{s-s_L} \sqrt{s-s_R}+m_N^2-m_{\pi }^2+3 s\right)\nonumber \\
    &-\ln \left(-\sqrt{s-s_L} \sqrt{s-s_R}+m_N^2-m_{\pi }^2+3 s\right)\ .
\end{align}
\section{CGLN Amplitudes}\label{ap:CGLN}
Traditional pion photoproduction partial wave analysis is in CGLN amplitudes (\(\mathcal{F}\)) with
\begin{equation}
    \frac{d\sigma}{d\Omega}=\frac{q'}{q}\left\lvert\langle\chi_f\lvert\mathcal{F}\rvert\chi_i\rangle\right\rvert^2\ ,
\end{equation}
where \(\chi_{i(f)}\) are Pauli spinor and 
\begin{align}
    \label{CGLN2}
    \mathcal{F}=i\vec{\sigma}\cdot\vec{\epsilon}\mathcal{F}_1+\left(\vec{\sigma}\cdot\vec{q'}\right)\vec{\sigma}\cdot\left(\vec{q}\times\vec{\epsilon}\right)\mathcal{F}_2
    +i\left(\vec{\sigma}\cdot q\right)\left(\vec{q'}\cdot\vec{\epsilon}\right)\mathcal{F}_3+i\left(\vec{q'}\cdot\vec{\sigma}\right)\left(\vec{q'}\cdot\vec{\epsilon}\right)\mathcal{F}_4\ ,
\end{align}
where there are four independent amplitudes.
The connection of our scattering amplitudes to \(\mathcal{F}\) can be obtained:
\begin{equation}
    \mathcal{M}_{fi}=8\pi\sqrt{s}\mathcal{F}_{fi}\ ,
\end{equation}
where the subscripts \(f,\ i\) mean initial and final states are substituted into Eq.~\ref{CGLN2} and we will omit it in the following discussion.

Furthermore, the partial wave amplitude \(\mathcal{F}^J\) is defined in Ref.~\cite{Review1969}:
\begin{equation}
    \mathcal{F}_{\pm;\lambda_r}^J=\frac{1}{4\pi}\int_{-1}^{1}\int_{0}^{2\pi} F_{\pm;\lambda_r}D^J(\theta,\phi){\rm d}\Omega\ ,
\end{equation}
where \(\pm \) mean the final nucleon helicity and \(\lambda_r=\frac{1}{2}\ \text{or}\ \frac{3}{2}\), which is the moduli of initial helicity. Also, definite parity amplitudes can be obtained:
\begin{align}
    A_{n+}=&-\frac{1}{\sqrt{2}}\left(\mathcal{F}_{+,\frac{1}{2}}^{J}+\mathcal{F}^{J}_{-,\frac{1}{2}}\right)\ ,\nonumber \\
    A_{\left(n+1\right)-}=&\frac{1}{\sqrt{2}}\left(\mathcal{F}_{+,\frac{1}{2}}^{J}-\mathcal{F}^{J}_{-,\frac{1}{2}}\right)\ ,\nonumber \\
    B_{n+}=&\sqrt{\frac{2}{n\left(n+2\right)}}\left(\mathcal{F}_{+,\frac{3}{2}}^{J}-\mathcal{F}^{J}_{-,\frac{3}{2}}\right)\ ,\\
    B_{\left(n+1\right)-}=&-\sqrt{\frac{2}{n\left(n+2\right)}}\left(\mathcal{F}_{+,\frac{3}{2}}^{J}-\mathcal{F}^{J}_{-,\frac{3}{2}}\right)\ ,\nonumber
\end{align}
where \(A_{n\pm},B_{n\pm}\) are amplitudes with \(J=n\pm\frac{1}{2}\) and \(P=-{\left(-1\right)}^n\).

According to Ref.~\cite{Review1969}, one can obtain the following relation between CGLN partial wave amplitudes (\(A_{n\pm}\) and \(B_{n\pm}\)) and multipole amplitudes (\(E_{n\pm}\) and \(M_{n\pm}\)):
\begin{align}
    E_{0+}=&A_{0+}\ ,\\
    M_{1-}=&A_{1-}\ .
\end{align}
and for \(l\geqslant 1\)
\begin{align}
    E_{l}+=&{(l+1)}^{-1}\left(A_{l+}+\frac{1}{2}lB_{l+}\right)\ ,\\
    M_{l}+=&{(l+1)}^{-1}\left(A_{l+}-\frac{1}{2}\left(l+2\right)B_{l+}\right)\ ,\\
    E_{(l+1)-}=&-{(l+1)}^{-1}\left(A_{(l+1)-}-\frac{1}{2}\left(l+2\right)B_{(l+1)-}\right)\ ,\\
    M_{(l+1)-}=&{(l+1)}^{-1}\left(A_{(l+1)-}+\frac{1}{2}lB_{(l+1)-}\right)\ .
\end{align}
Further, consider the fact that \(E^{I=\frac{1}{2}}_{0+}\) isn't normalized in isospin space according to Refs.~\cite{Amdt1990} and~\cite{Gasparyan2010}, so we have additional \(\sqrt{3}\) in normalization factor, and the relation in \(S_{11}\) channel can be obtained:
\begin{equation}\label{nbmp}
    E_{0+}^{I=\frac{1}{2}}=-\sqrt{\frac{2}{3s}}\mathcal{M}\left(S_{11}\right)\ ,
\end{equation}
where \(E_{0+}^{I=\frac{1}{2}}\) is conventional multipole amplitude with \(0\) and \(+\) refers to S wave and minus parity respectively.

\end{appendices}

\newpage

\bibliographystyle{h-physrev}
\bibliography{photoN}

\end{document}